\documentclass[%
 aps,
 prl,%
 amsmath,amssymb,
 twocolumn,%
]{revtex4-2}

\usepackage{graphicx}
\usepackage{dcolumn}
\usepackage{bm}
\usepackage{hyperref}
\usepackage{color}
\usepackage{appendix}
\usepackage{siunitx}
\usepackage{ulem}

\newcommand{\co}{\mathcal{C}}

\newcommand{\br}{{\bf r}}
\newcommand{\bx}{{\bf x}}

\newcommand{\avg}[1]{\langle #1 \rangle}

\begin{document}

\preprint{AIP/123-QED}

\title[]{Coherence of velocity fluctuations in turbulent flows}

\author{G. Prabhudesai}
\email{gaurav.prabhudesai@ens.fr}
\author{S. Perrard}
\author{F. P\'{e}tr\'{e}lis}
\author{S. Fauve}
\affiliation{Laboratoire de Physique de l'\'{E}cole Normale Sup\'{e}rieure, CNRS, PSL Research University, Sorbonne Universit\'{e}, Universit\'{e} de Paris, F-75005 Paris, France}

\date{\today}

\begin{abstract}
We investigate the spatio-temporal quantity of coherence for turbulent velocity fluctuations at spatial distances of the order or larger than the integral length scale $l_{0}$.  Using   controlled laboratory experiments, an exponential decay as a function of distance is observed with a decay rate which depends on the flow properties. The same law is observed in two different flows indicating that it can be a generic property of turbulent flows.  
\end{abstract}

\maketitle

\textit{Introduction---} Part of the spatial structure of turbulent flows has been extensively studied owing to the concept of energy cascade by Richardson~\cite{Richardson_book} and later extended to a scale invariant hypothesis in Kolmogorov's 1941 theory~\cite{Kolmogorov_1941}. In a three dimensional turbulent flow, a direct cascade of energy takes place between the integral length scale $l_{0}$ associated to the energy injection scale and the Kolmogorov length scale $\eta$ where energy gets eventually dissipated. In contrast, the spatio-temporal properties of velocity fluctuations in turbulent flows for separation distance $r$ comparable to or larger than the integral length scale $l_{0}$ has been significantly less studied.
Yet, understanding the behavior of turbulent fluctuations at large scales is not only of fundamental interest, it also has application for instance in geophysical or astrophysical flows when a large scale field bifurcates over a small scale turbulent flow \cite{bifturb}. Magnetic field generation by the alpha effect of astrophysical dynamos \cite{alpha}, large scale hydrodynamic flow generated by the AKA effect of helical flows \cite{aka} are two examples in which fluctuations at small scales may affect a large scale field, in particular if these fluctuations display some coherent behavior at large temporal or spatial scales.  The statistical properties of the large scales of turbulent flows are also of interest in industrial applications such as wind turbine farms for instance (see below). 
    

One tool to study the spatio-temporal velocity fluctuations at large scales in an homogeneous, stationary turbulent flow is the magnitude squared coherence or simply coherence defined from the signal at two points separated by vector \br~and a time lag $\tau$, as 
\begin{eqnarray}
    \mathcal{C}(\br,f) = \frac{|E_{ij}(\br,f)|^2}{E_{i}(f)E_{j}(f)}
\label{coh}
\end{eqnarray}
where $E_{ij}(\br,f) = \int_{-\infty}^{\infty} \langle u_{i}(\bx,t) u_{j}(\bx+\br,t+\tau) \rangle e^{\iota f \tau} d\tau$ is the cross-spectrum and $E_{i}(f) = \int_{-\infty}^{\infty} \langle u_{i}(\bx,t) u_{i}(\bx,t+\tau) \rangle e^{\iota f \tau} d\tau$ the one point spectrum of $i^{th}$ component. Coherence may refer to different velocity components. We denote in particular longitudinal (resp. transverse) the coherence function with $i=j$ and the velocity component parallel (resp. perpendicular) to ${\bf r}$. 

In the context of turbulent atmospheric boundary layer, a few  field experiments and modelling approaches have been devoted to the study of coherence~\cite{Davenport_1961,Thresher_1981,Kristensen_1979,Saranya_2004,Baker_2007,Krug_2019}, to estimate power load fluctuations in wind turbine farms~\cite{Vermeer_2003,Sorensen_2007} or to evaluate large scale constraints on bridges~\cite{Cheynet_2016} and buildings~\cite{Simiu_1996}. The coherence function is however still poorly documented, with measurements only in the context of turbulent atmospheric boundary layer and turbulent wakes. 

Despite $\mathcal{C}(\br,f)$ being a  standard quantity in signal analysis, it has been rarely used for turbulent data even though it  provides additional information about two-point correlation functions at equal time to which it is related but not in a simple way. 


In this letter, we investigate the behaviour of coherence in two controlled laboratory experiments, that we design to either vary the typical time scale or length scale of the flow. In both  set-ups, we also achieve a large separation of length scale between the size of the experimental domain and the integral length scale $l_{0}$.

\begin{figure}[htbp]
    \centering
\includegraphics[width=0.7\linewidth]{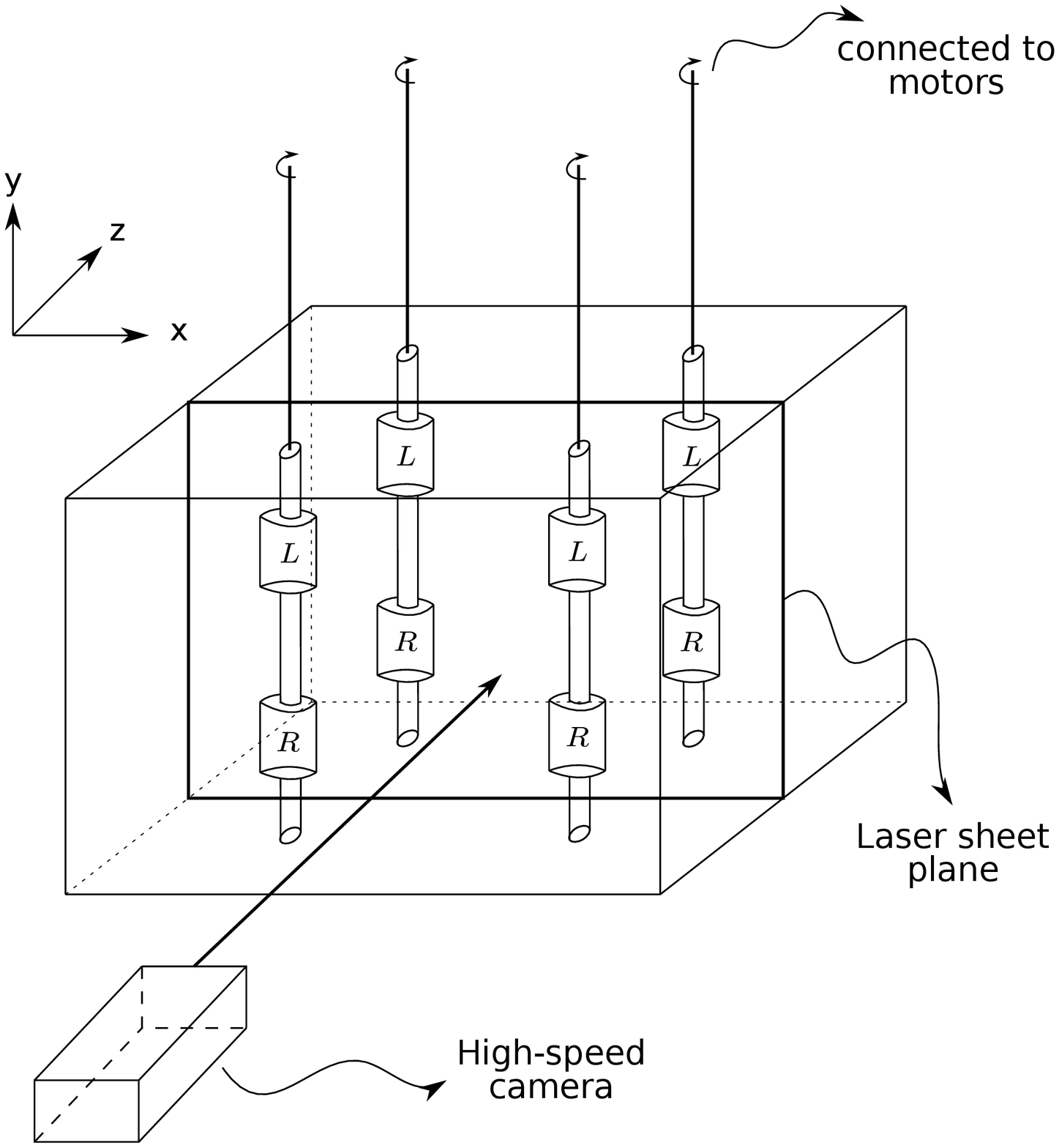}
\includegraphics[width=1\linewidth]{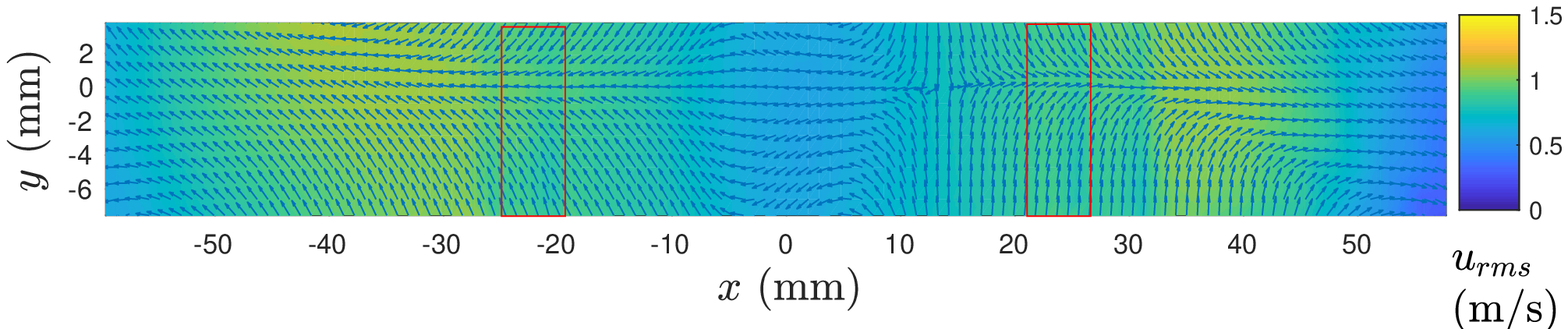}
\includegraphics[width=1\linewidth]{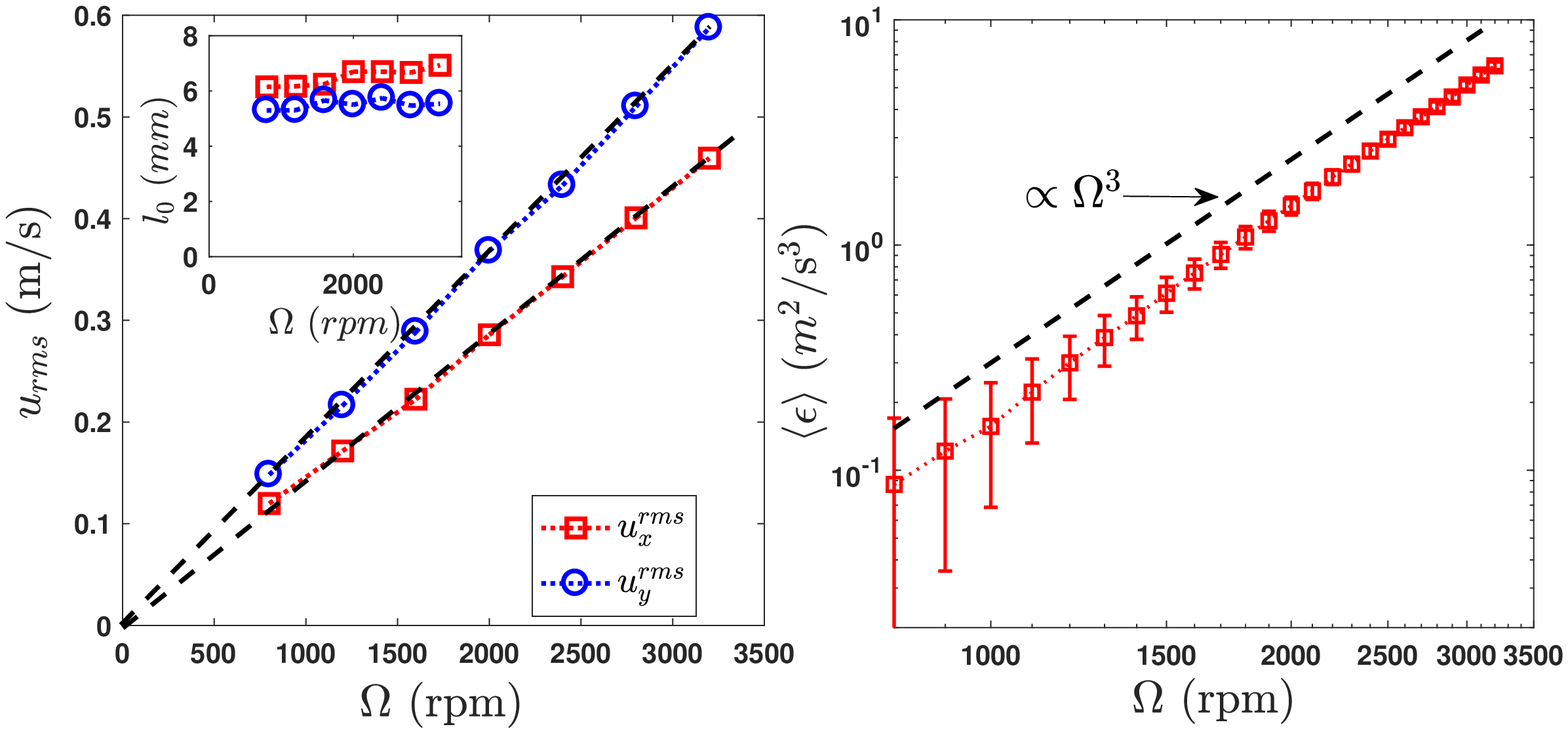}
\caption{(a) Sketch of the experimental set-up composed of 8 helices mounted on four vertical bars (R and L stand for right and left chirality). The velocity field is measured in the central $xy$ plane using PIV technique. (b) 2D map of the magnitude of velocity fluctuations $u_{rms}$. The blue arrows indicate the local direction of the mean flow, which evidence the stagnation point at the center. The position of the vertical bars are indicated in red. (c) RMS of velocity fluctuation along $x$ ($\textcolor{red}{\small{\Box}}$) and along $y$ ($\textcolor{blue}{\circ}$). Inset : longitudinal integral length scale $l_0$ along $x$ ($\textcolor{red}{\small{\Box}}$) and $y$ ($\textcolor{blue}{\circ}$).(d) Global mean energy injection rate per unit mass $\avg{\epsilon}$ evaluated from the motor power consumption.}
\label{fig:1}
\end{figure}

\textit{Experimental set-up and results---} The first experimental set-up is sketched in fig~\ref{fig:1}a. Four pairs of helices are mounted on vertical bars immersed in a cubical tank of side $15 \ \si{\centi\metre}$. Set in rotation by loop-controlled motors at an angular velocity $\Omega$, the helices force the flow directly in the bulk. The four axes are rotated in clockwise direction and $\Omega$ ranges from $0$ to $3200$ rpm. In order to obtain full optical access, we implement an index matching technique. We 3D print the helices in a transparent resin (Nano clear) of refractive index $n = 1.51$ and we match the resin refractive index by using a liquid mixture of $62\%$ in volume of anise oil (AO) and $38\%$ in volume of mineral oil (MO). As suggested by Song {\it et al.}~\cite{Song_2014}, the AO-MO mixture is particularly suitable for optical measurements of flows as a highly transparent, low viscosity ($\nu = 4.5$ cP for 62\% AO) fluid. We perform optical velocity measurements using both Laser Doppler Anemometry (LDA) and Particle Image Velocimetry (PIV). The LDA apparatus, composed of a pre-calibrated Dantec System, is used to characterize the local properties of the flow at small scale, while the PIV is used to study the large scale spatio-temporal fluctuations. The PIV is performed on vertical planes using a high speed camera and a continuous $2 \ \si{\watt}$ laser of wavelength $532 \ \si{\nano\metre}$. Using a combination of two spherical lenses and two cylindrical lenses, we obtain a vertical laser sheet with a tunable angle of divergence and a thickness of $500 \ \si{\micro\metre}$ in the tank. Particles of diameter $30 \pm 10 \ \si{\micro\metre}$ are seeded in the flow and the velocity field is reconstructed from the images using a free Particle Image Velocimetry algorithm~\cite{Thielicke_2014}. Fig.~\ref{fig:1}b shows the map of the two dimensional root mean squared velocity $u_{rms}^{tot} = \sqrt{{(u^{rms}_x)}^2 + {(u^{rms}_y)}^2}$ averaged over time. The blue arrow indicated the direction of the mean flow, showing the stagnation point at the center of the tank. This mean flow is however smaller than the velocity fluctuations in the region of interest.  The RMS of velocity fluctuations $u_{rms}$ along $x$ and $y$ vary linearly with the rotation rate (fig~\ref{fig:1}c). The correlation length of velocity fluctuations is computed using the two-point spatial correlation at equal time, which displays an exponential decay with a characteristic length defined as the integral length scale $l_{0}$. The integral length scale is measured to be independent of $\Omega$ (fig~\ref{fig:1}c inset) for the range of rotation rates studied  with $l_{0} = 6 \pm 1$~mm being significantly smaller than the box size. We measure the mean energy injection rate per unit mass $\langle \epsilon \rangle$ from the power required by the motors to maintain the flow. We find that $\langle \epsilon \rangle \propto \Omega^3$ (fig.~\ref{fig:1}d), which confirms that we reach the large scale-scaling at high Reynolds number $\langle \epsilon \rangle = C_{\epsilon} u_{rms}^3/l_0$  with the dimensionless constant $C_{\epsilon} = 0.4$ ~\cite{pope2001turbulent,vassilicos2015dissipation}. The rms of one velocity component is denoted by $u_{rms}$, whose scaling can also be recovered from dimensional arguments in the limit of high Reynolds number. Eventually, we estimate the Taylor Reynolds number $Re_\lambda = u_{rms} \lambda/\nu$ where $\lambda$ is the Taylor microscale associated to the correlation length of velocity gradients, which can be estimated from $u_{rms}$ and the dissipation rate, equal to the mean injection rate $\avg{\epsilon}$ in a statistically stationary regime. For the highest rotation speed, we achieve a Taylor Reynolds number $Re_\lambda \approx 100$ and $\lambda \approx 1$~mm. 

\begin{figure}[htbp]
    \centering
    \includegraphics[width=0.95\linewidth]{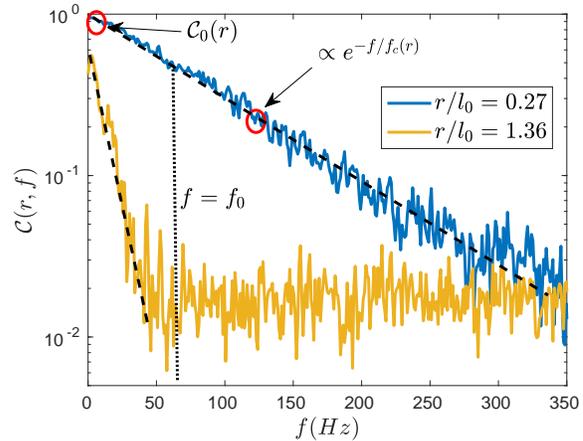}
\caption{Coherence as a function of frequency measured for $\Omega = 3200$ rpm at two spatial distances $r/l_{0} = 0.27$ (blue line) and $r/l_{0} = 1.36$ (yellow line).}
\label{fig:2}
\end{figure}

We focus on the large scale behavior corresponding to $r$ of the order or larger than $l_0$. Unless otherwise stated, measurements are reported for $r/l_0 \ge 0.27$.  Fig.~\ref{fig:2} shows coherence for the longitudinal component of the velocity as a function of frequency for two different values of distance $r$ at $\Omega = 3200$ rpm. We observe an exponential decay $\co = \co_0(r) exp(-f/f_{c})$ where $f_{c}^{-1}$ is the decay rate in frequency and $\co_0$ the coherence at zero frequency. Both $\co_0$ and $f_{c}$ are observed to decay with $r$. Their values are evaluated from the least root-mean-square error exponential fit of $\co$ from 1 to the noise level at large frequency. The decay of $\co_{0}$ with $r/l_0$ is shown in fig.~\ref{fig:3}a, each curve corresponding to a different value of $\Omega$. We find an exponential decay of $\mathcal{C}_0$ with a characteristic length scale proportional to the integral length scale $l_{0}$.  

These measurements imply that $\co_0 = \exp(-c_{1} r/l_{0})$. Further analysis on the frequency dependence of coherence reveals $f_{c} = c_{2} u_{rms}/r$ where $u_{rms}$ is the RMS of the longitudinal component of velocity. This can be seen in fig.~\ref{fig:3}b where coherence normalized by its value at zero frequency is plotted against normalized frequency $rf/u_{rms}$ for two different values of $r$ and of $\Omega$. The four curves lie within a single master curve. The experimentally observed behaviour of coherence is hence of the form
\begin{eqnarray}
    \mathcal{C}(r,f) =  \exp(-c_{1} r/l_0) \exp(-c_{2} r f/u_{rms})
\label{coh2}
\end{eqnarray}
with $c_1 = 0.54$ and $c_2 = 2.5$, valid for $r/l_0 \ge 0.27$. This functional form remains valid for the transverse component of velocity.


\begin{figure}[htbp]
    \centering
    \includegraphics[width=1\linewidth]{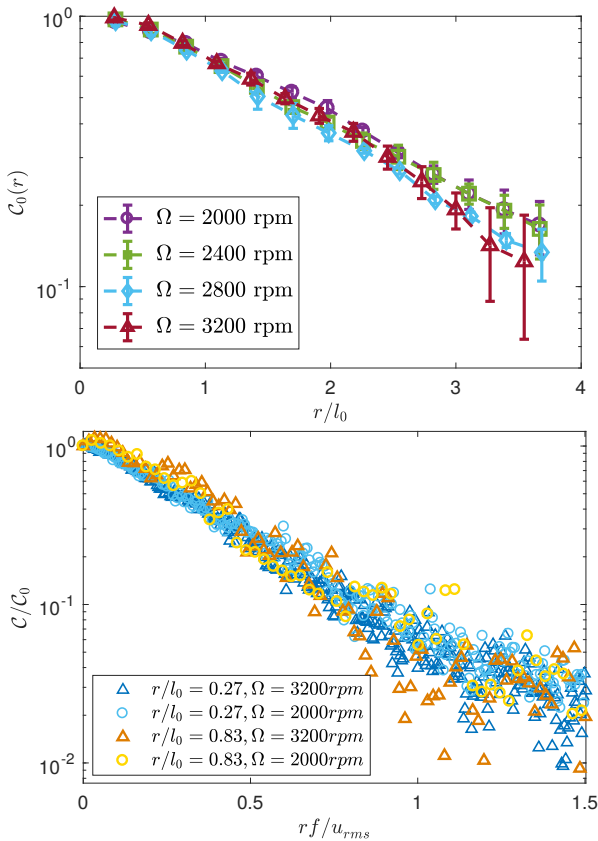}
    \label{fig:3}
\caption{(a) Coherence at zero frequency $\mathcal{C}_{0}$ as a function of the dimensionless distance $r/l_0$. (b) Normalized coherence $\mathcal{C}/\mathcal{C}_{0}$ as a function of the dimensionless frequency $rf/u_{rms}$.}
\label{fig:3}
\end{figure}

In this first experiment, however, the integral length scale does not vary significantly. To further check the relation $\co_0 = exp(-r/l_0)$, we design a second experiment where a turbulent flow is generated in air between two square walls (length $1$ m) of counter-rotating staggered fans. A sketch of the set-up is shown in fig.~\ref{fig:4}a. This configuration creates a Roberts-like turbulent flow~\cite{roberts1972dynamo} between the two walls. We use two 1D hot-wire probes using Constant Temperature Anemometry (CTA) technique to measure the two-point spatio-temporal quantities of speed fluctuations near the center. We vary the distance between the walls from $20$ cm to $80$ cm, which modifies the integral scale $l_{0}$ of the flow from $1.5$ cm to $3.5$ cm. Concomitantly, the RMS of velocity fluctuations $u_{rms}$ ranges from $0.1$ m/s to $0.5$ m/s. We get a maximum Taylor Reynolds number $Re_{\lambda} \approx 100$ for the minimum wall distance, corresponding to an integral length scale $l_{0} \sim O(1)$ cm. The coherence $\co$ in this set-up also exhibits an exponential decay of the form given by eq.~\eqref{coh2} as shown in fig.~\ref{fig:4}b with $c_1=0.84$ and $c_2 = 1.1$. The value of $c_1$ is found to be similar between the two experiments. The value of $c_2$ is likely under-estimated in the second experiment due to the hot-wire technique, which evaluates the speed of the sum of longitudinal and one transverse velocity component instead of the  longitudinal velocity. The two experimental configuration confirm the functional form of $\co$ of Eq.~\ref{coh2}, and in particular the dependency on the integral length scale $l_{0}$ that can here be varied. 

\begin{figure}[htbp]
    \centering
        \includegraphics[height=7cm,width=0.85\linewidth]{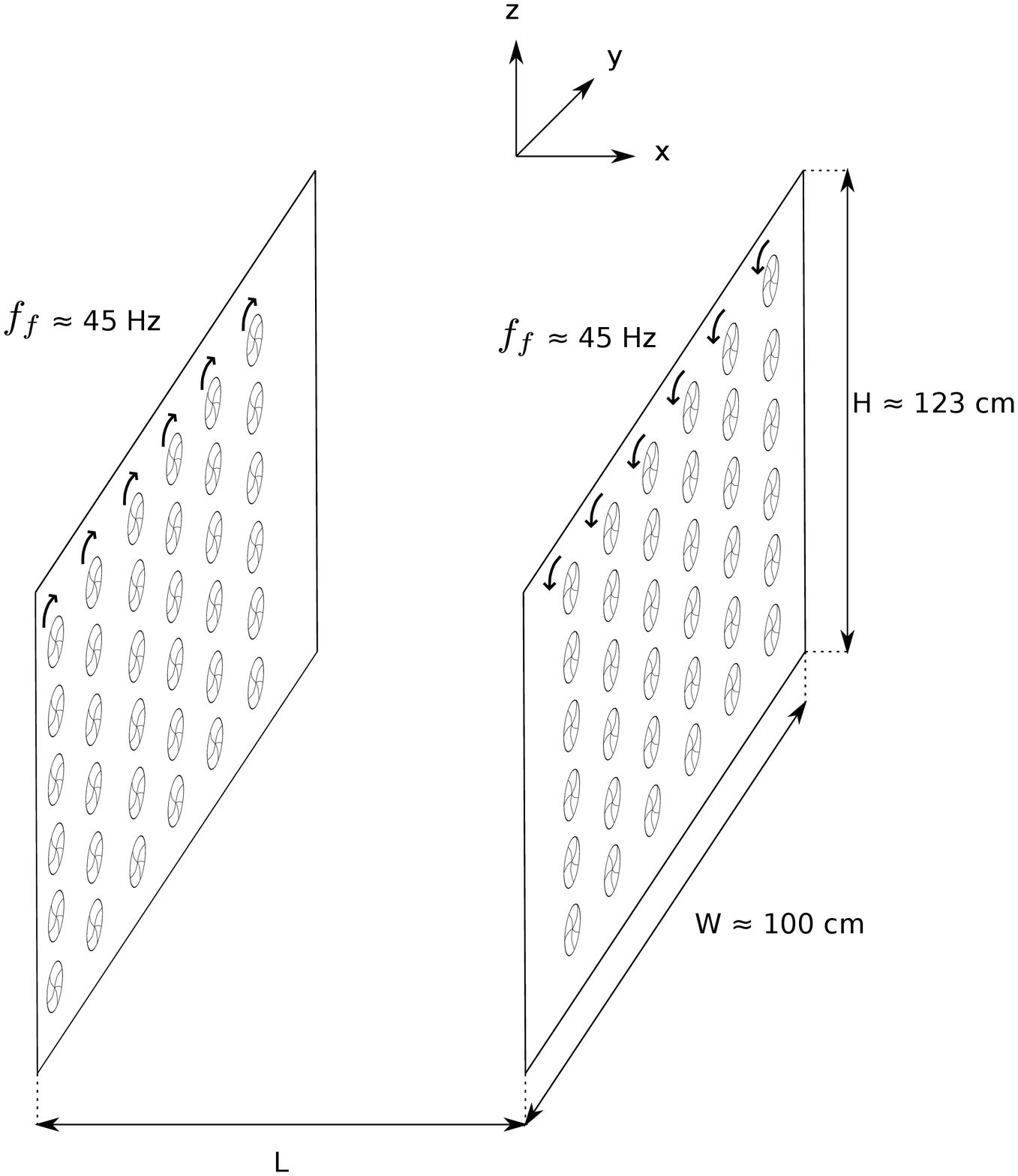}
        \includegraphics[width=1\linewidth]{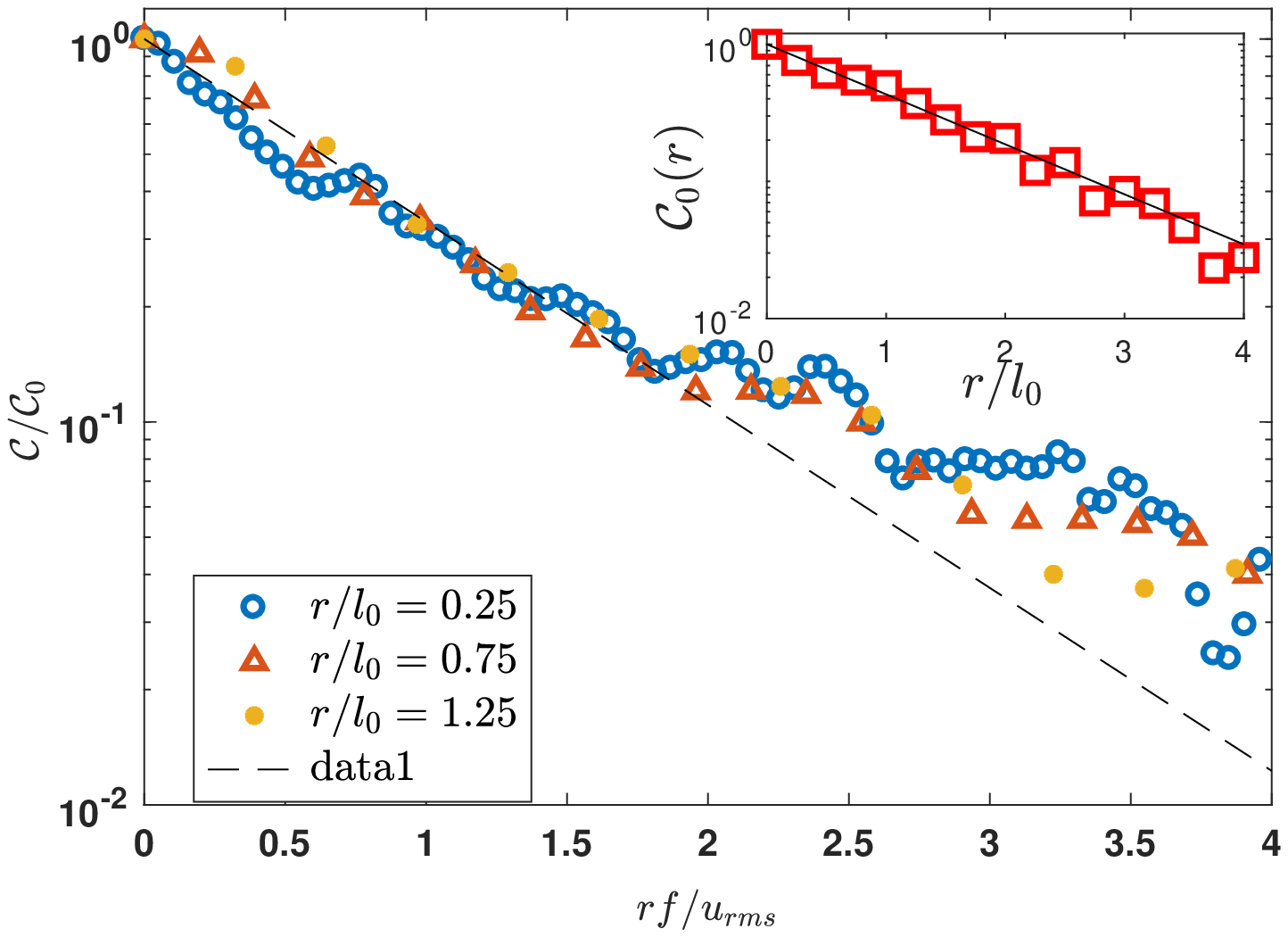}
        \label{fig:4}
\caption{a) Sketch of the experimental set-up of two walls of counter-rotating staggered fans. Each wall is fitted with 36 fans. b) Normalized coherence $C/C_0$ as a function of $r f/u_{rms}$. Inset : coherence $C_0$ at zero frequency as a function of the dimensionless distance $r/l_0$. $C/C_0$ and $C_0$ exhibit exponential decay resp. in $r f/u_{rms}$ and $r/l_0$.}
\label{fig:4}
\end{figure}

\textit{Physical interpretation---} In an homogeneous flow, coherence is an even function of $r$. Assuming that viscous effects render the flow properties smooth at small enough distances, coherence is expected to be quadratic in $r$ at small $r$. By definition it is equal to $1$ at $r=0$. 
Using the random sweeping hypothesis (RSH) as formulated by  Tennekes~\cite{Tennekes_1975} and Kraichnan~\cite{chen1989sweeping}, {Tobin \& Chamorro}~\cite{Tobin_2018} predicted  $\mathcal{C}(r,f) \propto \exp(-c r^2 f^2/U^2)$ in the case of a large mean flow $U$. 
The experiments that we report here are well designed to study the large $r$ behavior. Results at small $r$ are compatible with a quadratic behavior but even for $r/l_0$ of the order of $1/4$  we observe that the frequency dependence of coherence has the form $\mathcal{C}(r,f) \propto \exp(-c r f/u'_{rms})$ thus decaying slower than the prediction from RSH. 


An exponential decay of coherence can be interpreted as follows. In an homogeneous flow, the spatial dependence of coherence is set by the velocity cross-correlation. In the context of stochastic processes of ordinary differential equation, the exponential decay of the temporal correlation is associated to a memory-less (Markovian) process. A similar argument can be used  in the spatial domain. Measuring distances along a line,  we assume that the coherence between $0$ and $ r+\Delta$ is equal  to the coherence between $0$ and $ r$ multiplied by a function that only depends on $\Delta$. We then have $C(r+\Delta,f)=C(r,f)\,g(\Delta)$. This functional relation leads to an exponential decay $C(r,f)\propto \exp{(-\Gamma r)}$ where $\Gamma^{-1}$ is the spatial correlation length. Our measurements show that 
\begin{eqnarray}
   \Gamma = \frac{c_{1}  f}{u'_{rms}}+\frac{c_{2}}{l_{0}}
\label{defGamma}
\end{eqnarray}
This exponential behavior of the coherence takes place at separation distance $r$ not too small, {\it i.e.} when the sweeping effect of the turbulent fluctuations has lost its coherence. The form of the correlation length $\Gamma^{-1}$ interpolates between the injection length scale at small $f$ and $u'_{rms}/f$ at large $f$.

We note that among the different empirical models introduced for coherence in turbulent atmospheric boundary layers, an exponential decay in the vertical coordinate is usually considered with a prefactor that depends on frequency. This is for instance the case of Davenport model as later refined by Thresher et al.  $\mathcal{C}(z,f) \propto \exp(-\sqrt{a (z f/U)^2+(b z/L)^2})$~\cite{Davenport_1961,Thresher_1981,Saranya_2004}, 
with $U$ the horizontal mean flow velocity,  $L$ a characteristic length, and $a$ and $b$ numerical constants. The asymptotic behavior at small or large $f$ are similar to our results of eq.~\eqref{coh2} even though the contexts are different: specific role played by the vertical coordinate, presence of a strong mean flow that renders Taylor' hypothesis valid and the consideration of only the inertial scales. 

\textit{Conclusion---}
We have shown experimentally that at scales larger than a fraction of the integral scale (for $r/l_0 \ge 0.27$), the coherence of the velocity in a turbulent flow decays exponentially in frequency and in space with a decay rate of the form of eq.~\ref{defGamma}. Being observed in two  different experimental set-ups, we believe that our observations are generic to the large scales of  turbulent flows.

\vspace*{5mm}

\begin{acknowledgments}
This work has been supported by the Agence nationale de la recherche (Grant No. ANR-17-CE30-0004) and CEFIPRA (Project No. 6104-1). 
\end{acknowledgments}

\nocite{*}
\bibliography{aipsamp}

\providecommand{\noopsort}[1]{}\providecommand{\singleletter}[1]{#1}%
\begin{thebibliography}{23}%
\makeatletter
\providecommand \@ifxundefined [1]{%
 \@ifx{#1\undefined}
}%
\providecommand \@ifnum [1]{%
 \ifnum #1\expandafter \@firstoftwo
 \else \expandafter \@secondoftwo
 \fi
}%
\providecommand \@ifx [1]{%
 \ifx #1\expandafter \@firstoftwo
 \else \expandafter \@secondoftwo
 \fi
}%
\providecommand \natexlab [1]{#1}%
\providecommand \enquote  [1]{``#1''}%
\providecommand \bibnamefont  [1]{#1}%
\providecommand \bibfnamefont [1]{#1}%
\providecommand \citenamefont [1]{#1}%
\providecommand \href@noop [0]{\@secondoftwo}%
\providecommand \href [0]{\begingroup \@sanitize@url \@href}%
\providecommand \@href[1]{\@@startlink{#1}\@@href}%
\providecommand \@@href[1]{\endgroup#1\@@endlink}%
\providecommand \@sanitize@url [0]{\catcode `\\12\catcode `\$12\catcode
  `\&12\catcode `\#12\catcode `\^12\catcode `\_12\catcode `\%12\relax}%
\providecommand \@@startlink[1]{}%
\providecommand \@@endlink[0]{}%
\providecommand \url  [0]{\begingroup\@sanitize@url \@url }%
\providecommand \@url [1]{\endgroup\@href {#1}{\urlprefix }}%
\providecommand \urlprefix  [0]{URL }%
\providecommand \Eprint [0]{\href }%
\providecommand \doibase [0]{https://doi.org/}%
\providecommand \selectlanguage [0]{\@gobble}%
\providecommand \bibinfo  [0]{\@secondoftwo}%
\providecommand \bibfield  [0]{\@secondoftwo}%
\providecommand \translation [1]{[#1]}%
\providecommand \BibitemOpen [0]{}%
\providecommand \bibitemStop [0]{}%
\providecommand \bibitemNoStop [0]{.\EOS\space}%
\providecommand \EOS [0]{\spacefactor3000\relax}%
\providecommand \BibitemShut  [1]{\csname bibitem#1\endcsname}%
\let\auto@bib@innerbib\@empty
\bibitem [{\citenamefont {Richardson}(1922)}]{Richardson_book}%
  \BibitemOpen
  \bibfield  {author} {\bibinfo {author} {\bibfnamefont {L.~F.}\ \bibnamefont
  {Richardson}},\ }\href@noop {} {\emph {\bibinfo {title} {Weather prediction
  by numerical process}}}\ (\bibinfo  {publisher} {Cambridge University
  Press},\ \bibinfo {year} {1922})\BibitemShut {NoStop}%
\bibitem [{\citenamefont {Kolmogorov}(1941)}]{Kolmogorov_1941}%
  \BibitemOpen
  \bibfield  {author} {\bibinfo {author} {\bibfnamefont {A.~N.}\ \bibnamefont
  {Kolmogorov}},\ }\bibfield  {title} {\bibinfo {title} {The local structure of
  turbulence in incompressible viscous fluid for very large reynolds numbers},\
  }\href@noop {} {\bibfield  {journal} {\bibinfo  {journal} {Dokl. Akad. Nauk}\
  }\textbf {\bibinfo {volume} {SSSR 30:301}} (\bibinfo {year}
  {1941})}\BibitemShut {NoStop}%
\bibitem [{\citenamefont {Fauve}\ \emph {et~al.}(2017)\citenamefont {Fauve},
  \citenamefont {Herault}, \citenamefont {Michel},\ and\ \citenamefont
  {P{\'{e}}tr{\'{e}}lis}}]{bifturb}%
  \BibitemOpen
  \bibfield  {author} {\bibinfo {author} {\bibfnamefont {S.}~\bibnamefont
  {Fauve}}, \bibinfo {author} {\bibfnamefont {J.}~\bibnamefont {Herault}},
  \bibinfo {author} {\bibfnamefont {G.}~\bibnamefont {Michel}},\ and\ \bibinfo
  {author} {\bibfnamefont {F.}~\bibnamefont {P{\'{e}}tr{\'{e}}lis}},\
  }\bibfield  {title} {\bibinfo {title} {Instabilities on a turbulent
  background},\ }\href@noop {} {\bibfield  {journal} {\bibinfo  {journal}
  {Journal of Statistical Mechanics: Theory and Experiment}\ }\textbf {\bibinfo
  {volume} {2017}},\ \bibinfo {pages} {064001} (\bibinfo {year}
  {2017})}\BibitemShut {NoStop}%
\bibitem [{\citenamefont {Moffatt}(1978)}]{alpha}%
  \BibitemOpen
  \bibfield  {author} {\bibinfo {author} {\bibfnamefont {K.}~\bibnamefont
  {Moffatt}},\ }\href@noop {} {\emph {\bibinfo {title} {The Generation of
  Magnetic Fields in Electrically Conducting Fluids}}}\ (\bibinfo {year}
  {1978})\BibitemShut {NoStop}%
\bibitem [{\citenamefont {Frisch}\ \emph {et~al.}(1987)\citenamefont {Frisch},
  \citenamefont {She},\ and\ \citenamefont {Sulem}}]{aka}%
  \BibitemOpen
  \bibfield  {author} {\bibinfo {author} {\bibfnamefont {U.}~\bibnamefont
  {Frisch}}, \bibinfo {author} {\bibfnamefont {Z.}~\bibnamefont {She}},\ and\
  \bibinfo {author} {\bibfnamefont {P.}~\bibnamefont {Sulem}},\ }\bibfield
  {title} {\bibinfo {title} {Large-scale flow driven by the anisotropic kinetic
  alpha effect},\ }\href
  {https://doi.org/https://doi.org/10.1016/0167-2789(87)90026-1} {\bibfield
  {journal} {\bibinfo  {journal} {Physica D: Nonlinear Phenomena}\ }\textbf
  {\bibinfo {volume} {28}},\ \bibinfo {pages} {382} (\bibinfo {year}
  {1987})}\BibitemShut {NoStop}%
\bibitem [{\citenamefont {Davenport}(1960)}]{Davenport_1961}%
  \BibitemOpen
  \bibfield  {author} {\bibinfo {author} {\bibfnamefont {A.~G.}\ \bibnamefont
  {Davenport}},\ }\bibfield  {title} {\bibinfo {title} {The spectrum of
  horizontal gustiness near the ground in high winds},\ }\href@noop {}
  {\bibfield  {journal} {\bibinfo  {journal} {Q. J. R. Meteorol. Soc.}\
  }\textbf {\bibinfo {volume} {87}},\ \bibinfo {pages} {194} (\bibinfo {year}
  {1960})}\BibitemShut {NoStop}%
\bibitem [{\citenamefont {Thresher}\ \emph {et~al.}(1981)\citenamefont
  {Thresher}, \citenamefont {Holley}, \citenamefont {Smith}, \citenamefont
  {Jafarey},\ and\ \citenamefont {Lin}}]{Thresher_1981}%
  \BibitemOpen
  \bibfield  {author} {\bibinfo {author} {\bibfnamefont {R.~W.}\ \bibnamefont
  {Thresher}}, \bibinfo {author} {\bibfnamefont {W.~E.}\ \bibnamefont
  {Holley}}, \bibinfo {author} {\bibfnamefont {C.~E.}\ \bibnamefont {Smith}},
  \bibinfo {author} {\bibfnamefont {N.}~\bibnamefont {Jafarey}},\ and\ \bibinfo
  {author} {\bibfnamefont {S.-R.}\ \bibnamefont {Lin}},\ }\href@noop {} {\emph
  {\bibinfo {title} {Modeling the response of wind turbines to atmospheric
  turbulence}}},\ \bibinfo {type} {Tech. Rep.}\ \bibinfo {number} {No.
  RL0/2227-81/2}\ (\bibinfo  {institution} {Department of Mechanical
  Engineering, Oregon State University},\ \bibinfo {year} {1981})\BibitemShut
  {NoStop}%
\bibitem [{\citenamefont {Kristensen}\ and\ \citenamefont
  {Jensen}(1979)}]{Kristensen_1979}%
  \BibitemOpen
  \bibfield  {author} {\bibinfo {author} {\bibfnamefont {L.}~\bibnamefont
  {Kristensen}}\ and\ \bibinfo {author} {\bibfnamefont {N.~O.}\ \bibnamefont
  {Jensen}},\ }\bibfield  {title} {\bibinfo {title} {Lateral coherence in
  isotropic turbulence and in the natural wind},\ }\href@noop {} {\bibfield
  {journal} {\bibinfo  {journal} {Boundary-Layer Met.}\ }\textbf {\bibinfo
  {volume} {17}},\ \bibinfo {pages} {353} (\bibinfo {year} {1979})}\BibitemShut
  {NoStop}%
\bibitem [{\citenamefont {Saranyasoontorn}\ \emph {et~al.}(2004)\citenamefont
  {Saranyasoontorn}, \citenamefont {Manuel},\ and\ \citenamefont
  {Veers}}]{Saranya_2004}%
  \BibitemOpen
  \bibfield  {author} {\bibinfo {author} {\bibfnamefont {K.}~\bibnamefont
  {Saranyasoontorn}}, \bibinfo {author} {\bibfnamefont {L.}~\bibnamefont
  {Manuel}},\ and\ \bibinfo {author} {\bibfnamefont {P.~S.}\ \bibnamefont
  {Veers}},\ }\bibfield  {title} {\bibinfo {title} {A comparison of standard
  coherence models for inflow turbulence with estimates from field
  measurements},\ }\href@noop {} {\bibfield  {journal} {\bibinfo  {journal} {J.
  of Solar Energy Eng.}\ }\textbf {\bibinfo {volume} {126}},\ \bibinfo {pages}
  {1069} (\bibinfo {year} {2004})}\BibitemShut {NoStop}%
\bibitem [{\citenamefont {Baker}(2007)}]{Baker_2007}%
  \BibitemOpen
  \bibfield  {author} {\bibinfo {author} {\bibfnamefont {C.~J.}\ \bibnamefont
  {Baker}},\ }\bibfield  {title} {\bibinfo {title} {Wind engineering: past,
  present and future},\ }\href@noop {} {\bibfield  {journal} {\bibinfo
  {journal} {J. Wind Eng. Ind. Aero.}\ }\textbf {\bibinfo {volume} {95}},\
  \bibinfo {pages} {843} (\bibinfo {year} {2007})}\BibitemShut {NoStop}%
\bibitem [{\citenamefont {Krug}\ \emph {et~al.}(2019)\citenamefont {Krug},
  \citenamefont {Baars}, \citenamefont {Hutchins},\ and\ \citenamefont
  {Marusic}}]{Krug_2019}%
  \BibitemOpen
  \bibfield  {author} {\bibinfo {author} {\bibfnamefont {D.}~\bibnamefont
  {Krug}}, \bibinfo {author} {\bibfnamefont {W.~J.}\ \bibnamefont {Baars}},
  \bibinfo {author} {\bibfnamefont {N.}~\bibnamefont {Hutchins}},\ and\
  \bibinfo {author} {\bibfnamefont {I.}~\bibnamefont {Marusic}},\ }\bibfield
  {title} {\bibinfo {title} {Vertical coherence of turbulence in the
  atmospheric surface layer: connecting the hypotheses of townsend and
  davenport},\ }\href@noop {} {\bibfield  {journal} {\bibinfo  {journal}
  {Boundary-Layer Meteorology}\ }\textbf {\bibinfo {volume} {172}},\ \bibinfo
  {pages} {199} (\bibinfo {year} {2019})}\BibitemShut {NoStop}%
\bibitem [{\citenamefont {Vermeer}\ \emph {et~al.}(2003)\citenamefont
  {Vermeer}, \citenamefont {Sorensen},\ and\ \citenamefont
  {Crespo}}]{Vermeer_2003}%
  \BibitemOpen
  \bibfield  {author} {\bibinfo {author} {\bibfnamefont {L.~J.}\ \bibnamefont
  {Vermeer}}, \bibinfo {author} {\bibfnamefont {J.~N.}\ \bibnamefont
  {Sorensen}},\ and\ \bibinfo {author} {\bibfnamefont {A.}~\bibnamefont
  {Crespo}},\ }\bibfield  {title} {\bibinfo {title} {Wind turbine wake
  aerodynamics},\ }\href@noop {} {\bibfield  {journal} {\bibinfo  {journal}
  {Prog. in Aerosp. Sci.}\ }\textbf {\bibinfo {volume} {6}},\ \bibinfo {pages}
  {467} (\bibinfo {year} {2003})}\BibitemShut {NoStop}%
\bibitem [{\citenamefont {Sorensen}\ \emph {et~al.}(2007)\citenamefont
  {Sorensen}, \citenamefont {Cutululis}, \citenamefont {Vigueras-Rodriguez},
  \citenamefont {Madsen}, \citenamefont {Pinson}, \citenamefont {Jensen},
  \citenamefont {Hjerrild},\ and\ \citenamefont {Donovan}}]{Sorensen_2007}%
  \BibitemOpen
  \bibfield  {author} {\bibinfo {author} {\bibfnamefont {P.}~\bibnamefont
  {Sorensen}}, \bibinfo {author} {\bibfnamefont {N.~A.}\ \bibnamefont
  {Cutululis}}, \bibinfo {author} {\bibfnamefont {A.}~\bibnamefont
  {Vigueras-Rodriguez}}, \bibinfo {author} {\bibfnamefont {H.}~\bibnamefont
  {Madsen}}, \bibinfo {author} {\bibfnamefont {P.}~\bibnamefont {Pinson}},
  \bibinfo {author} {\bibfnamefont {L.~E.}\ \bibnamefont {Jensen}}, \bibinfo
  {author} {\bibfnamefont {J.}~\bibnamefont {Hjerrild}},\ and\ \bibinfo
  {author} {\bibfnamefont {M.}~\bibnamefont {Donovan}},\ }\bibfield  {title}
  {\bibinfo {title} {Power fluctuations from large wind farms},\ }\href@noop {}
  {\bibfield  {journal} {\bibinfo  {journal} {IEEE Trans. Power Syst.}\
  }\textbf {\bibinfo {volume} {22}},\ \bibinfo {pages} {958} (\bibinfo {year}
  {2007})}\BibitemShut {NoStop}%
\bibitem [{\citenamefont {Cheynet}\ \emph {et~al.}(2016)\citenamefont
  {Cheynet}, \citenamefont {Jakobsen},\ and\ \citenamefont
  {Snaebjornsson}}]{Cheynet_2016}%
  \BibitemOpen
  \bibfield  {author} {\bibinfo {author} {\bibfnamefont {E.}~\bibnamefont
  {Cheynet}}, \bibinfo {author} {\bibfnamefont {J.~B.}\ \bibnamefont
  {Jakobsen}},\ and\ \bibinfo {author} {\bibfnamefont {J.}~\bibnamefont
  {Snaebjornsson}},\ }\bibfield  {title} {\bibinfo {title} {Buffeting response
  of a suspension bridge in complex terrain},\ }\href@noop {} {\bibfield
  {journal} {\bibinfo  {journal} {Eng. structures}\ }\textbf {\bibinfo {volume}
  {128}},\ \bibinfo {pages} {474} (\bibinfo {year} {2016})}\BibitemShut
  {NoStop}%
\bibitem [{\citenamefont {Simiu}\ and\ \citenamefont
  {Scanlan}(1996)}]{Simiu_1996}%
  \BibitemOpen
  \bibfield  {author} {\bibinfo {author} {\bibfnamefont {E.}~\bibnamefont
  {Simiu}}\ and\ \bibinfo {author} {\bibfnamefont {R.}~\bibnamefont
  {Scanlan}},\ }\href@noop {} {\emph {\bibinfo {title} {Wind effects on
  structures: fundamentals and applications to design}}},\ \bibinfo {edition}
  {3rd}\ ed.\ (\bibinfo  {publisher} {Wiley, New York},\ \bibinfo {year}
  {1996})\BibitemShut {NoStop}%
\bibitem [{\citenamefont {Song}\ \emph {et~al.}(2015)\citenamefont {Song},
  \citenamefont {Choi}, \citenamefont {Seong},\ and\ \citenamefont
  {Kim}}]{Song_2014}%
  \BibitemOpen
  \bibfield  {author} {\bibinfo {author} {\bibfnamefont {M.~S.}\ \bibnamefont
  {Song}}, \bibinfo {author} {\bibfnamefont {H.~Y.}\ \bibnamefont {Choi}},
  \bibinfo {author} {\bibfnamefont {J.~H.}\ \bibnamefont {Seong}},\ and\
  \bibinfo {author} {\bibfnamefont {E.~S.}\ \bibnamefont {Kim}},\ }\bibfield
  {title} {\bibinfo {title} {Matching-index-of-refraction of transparent 3d
  printing models for flow visualization},\ }\href@noop {} {\bibfield
  {journal} {\bibinfo  {journal} {Nucl. Eng. and Design}\ }\textbf {\bibinfo
  {volume} {284}},\ \bibinfo {pages} {185} (\bibinfo {year}
  {2015})}\BibitemShut {NoStop}%
\bibitem [{\citenamefont {Thielicke}\ and\ \citenamefont
  {Stamhuis}(2014)}]{Thielicke_2014}%
  \BibitemOpen
  \bibfield  {author} {\bibinfo {author} {\bibfnamefont {W.}~\bibnamefont
  {Thielicke}}\ and\ \bibinfo {author} {\bibfnamefont {E.}~\bibnamefont
  {Stamhuis}},\ }\bibfield  {title} {\bibinfo {title} {Pivlab--towards
  user-friendly, affordable and accurate digital particle image velocimetry in
  matlab},\ }\href@noop {} {\bibfield  {journal} {\bibinfo  {journal} {Journal
  of open research software}\ }\textbf {\bibinfo {volume} {2}} (\bibinfo {year}
  {2014})}\BibitemShut {NoStop}%
\bibitem [{\citenamefont {Pope}(2001)}]{pope2001turbulent}%
  \BibitemOpen
  \bibfield  {author} {\bibinfo {author} {\bibfnamefont {S.~B.}\ \bibnamefont
  {Pope}},\ }\href@noop {} {\bibinfo {title} {Turbulent flows}} (\bibinfo
  {year} {2001})\BibitemShut {NoStop}%
\bibitem [{\citenamefont {Vassilicos}(2015)}]{vassilicos2015dissipation}%
  \BibitemOpen
  \bibfield  {author} {\bibinfo {author} {\bibfnamefont {J.~C.}\ \bibnamefont
  {Vassilicos}},\ }\bibfield  {title} {\bibinfo {title} {Dissipation in
  turbulent flows},\ }\href@noop {} {\bibfield  {journal} {\bibinfo  {journal}
  {Annual Review of Fluid Mechanics}\ }\textbf {\bibinfo {volume} {47}},\
  \bibinfo {pages} {95} (\bibinfo {year} {2015})}\BibitemShut {NoStop}%
\bibitem [{\citenamefont {Roberts}(1972)}]{roberts1972dynamo}%
  \BibitemOpen
  \bibfield  {author} {\bibinfo {author} {\bibfnamefont {G.~O.}\ \bibnamefont
  {Roberts}},\ }\bibfield  {title} {\bibinfo {title} {Dynamo action of fluid
  motions with two-dimensional periodicity},\ }\href@noop {} {\bibfield
  {journal} {\bibinfo  {journal} {Philosophical Transactions of the Royal
  Society of London. Series A, Mathematical and Physical Sciences}\ }\textbf
  {\bibinfo {volume} {271}},\ \bibinfo {pages} {411} (\bibinfo {year}
  {1972})}\BibitemShut {NoStop}%
\bibitem [{\citenamefont {Tennekes}(1975)}]{Tennekes_1975}%
  \BibitemOpen
  \bibfield  {author} {\bibinfo {author} {\bibfnamefont {H.}~\bibnamefont
  {Tennekes}},\ }\bibfield  {title} {\bibinfo {title} {Eulerian and lagrangian
  time microscales in isotropic turbulence},\ }\href@noop {} {\bibfield
  {journal} {\bibinfo  {journal} {J. Fluid Mech.}\ }\textbf {\bibinfo {volume}
  {67}},\ \bibinfo {pages} {561} (\bibinfo {year} {1975})}\BibitemShut
  {NoStop}%
\bibitem [{\citenamefont {Chen}\ and\ \citenamefont
  {Kraichnan}(1989)}]{chen1989sweeping}%
  \BibitemOpen
  \bibfield  {author} {\bibinfo {author} {\bibfnamefont {S.}~\bibnamefont
  {Chen}}\ and\ \bibinfo {author} {\bibfnamefont {R.~H.}\ \bibnamefont
  {Kraichnan}},\ }\bibfield  {title} {\bibinfo {title} {Sweeping decorrelation
  in isotropic turbulence},\ }\href@noop {} {\bibfield  {journal} {\bibinfo
  {journal} {Physics of Fluids A: Fluid Dynamics}\ }\textbf {\bibinfo {volume}
  {1}},\ \bibinfo {pages} {2019} (\bibinfo {year} {1989})}\BibitemShut
  {NoStop}%
\bibitem [{\citenamefont {Tobin}\ and\ \citenamefont
  {Chamorro}(2018)}]{Tobin_2018}%
  \BibitemOpen
  \bibfield  {author} {\bibinfo {author} {\bibfnamefont {N.}~\bibnamefont
  {Tobin}}\ and\ \bibinfo {author} {\bibfnamefont {L.~P.}\ \bibnamefont
  {Chamorro}},\ }\bibfield  {title} {\bibinfo {title} {Turbulence coherence and
  its impact on wind-farm power fluctuations},\ }\href@noop {} {\bibfield
  {journal} {\bibinfo  {journal} {J. Fluid Mech.}\ }\textbf {\bibinfo {volume}
  {855}},\ \bibinfo {pages} {1116} (\bibinfo {year} {2018})}\BibitemShut
  {NoStop}%
\end{thebibliography}%

\end{document}